\pdfoutput=1

\documentclass[english,aps,prb, superscriptaddress,twocolumn]{revtex4}
\usepackage{berasans}
\usepackage[latin9]{inputenc}
\usepackage{amssymb}
\usepackage{graphicx}
\usepackage{esint}

\makeatletter
\@ifundefined{textcolor}{}
{%
 \definecolor{BLACK}{gray}{0}
 \definecolor{WHITE}{gray}{1}
 \definecolor{RED}{rgb}{1,0,0}
 \definecolor{GREEN}{rgb}{0,1,0}
 \definecolor{BLUE}{rgb}{0,0,1}
 \definecolor{CYAN}{cmyk}{1,0,0,0}
 \definecolor{MAGENTA}{cmyk}{0,1,0,0}
 \definecolor{YELLOW}{cmyk}{0,0,1,0}
}

\usepackage{upgreek}

\newcommand{\II}{\mathrm{i}}

\makeatother

\usepackage{babel}
\begin{document}

\title{Macroscopic electromagnetic response of metamaterials with toroidal
resonances}

\author{V. Savinov}

\email{vs1106@orc.soton.ac.uk}

\affiliation{Optoelectronics Research Centre \& Centre for Photonic Metamaterials,
University of Southampton, UK}

\author{V. A. Fedotov}

\affiliation{Optoelectronics Research Centre \& Centre for Photonic Metamaterials,
University of Southampton, UK}

\author{N. I. Zheludev}

\affiliation{Optoelectronics Research Centre \& Centre for Photonic Metamaterials,
University of Southampton, UK}

\affiliation{Centre for Disruptive Photonic Technologies, Nanyang Technological
University, Singapore 637371, Singapore}
\begin{abstract}
Toroidal dipole, first described by Ia. B. Zeldovich {[}Sov. Phys.
JETP \textbf{33}, 1184 (1957){]}, is a distinct electromagnetic excitation
that differs both from the electric and the magnetic dipoles. It has
a number of intriguing properties: static toroidal nuclear dipole
is responsible for parity violation in atomic spectra; interactions
between static toroidal dipole and oscillating magnetic dipole are
claimed to violate Newton\textquoteright{}s Third Law while non-stationary
charge-current configurations involving toroidal multipoles have been
predicted to produce vector potential in the absence of electromagnetic
fields. Existence of the toroidal response in metamaterials was recently
demonstrated and is now a growing field of research. However, no direct
analytical link has yet been established between the transmission
and reflection of macroscopic electromagnetic media and toroidal dipole
excitations. To address this essential gap in electromagnetic theory
we have developed an analytical approach linking microscopic and macroscopic
electromagnetic response of a metamaterial and showed, using a case
study, the key role of the toroidal dipole in shaping the electromagnetic
properties of the metamaterial. 
\end{abstract}
\maketitle
\onecolumngrid
\newpage

\begin{figure}
\includegraphics{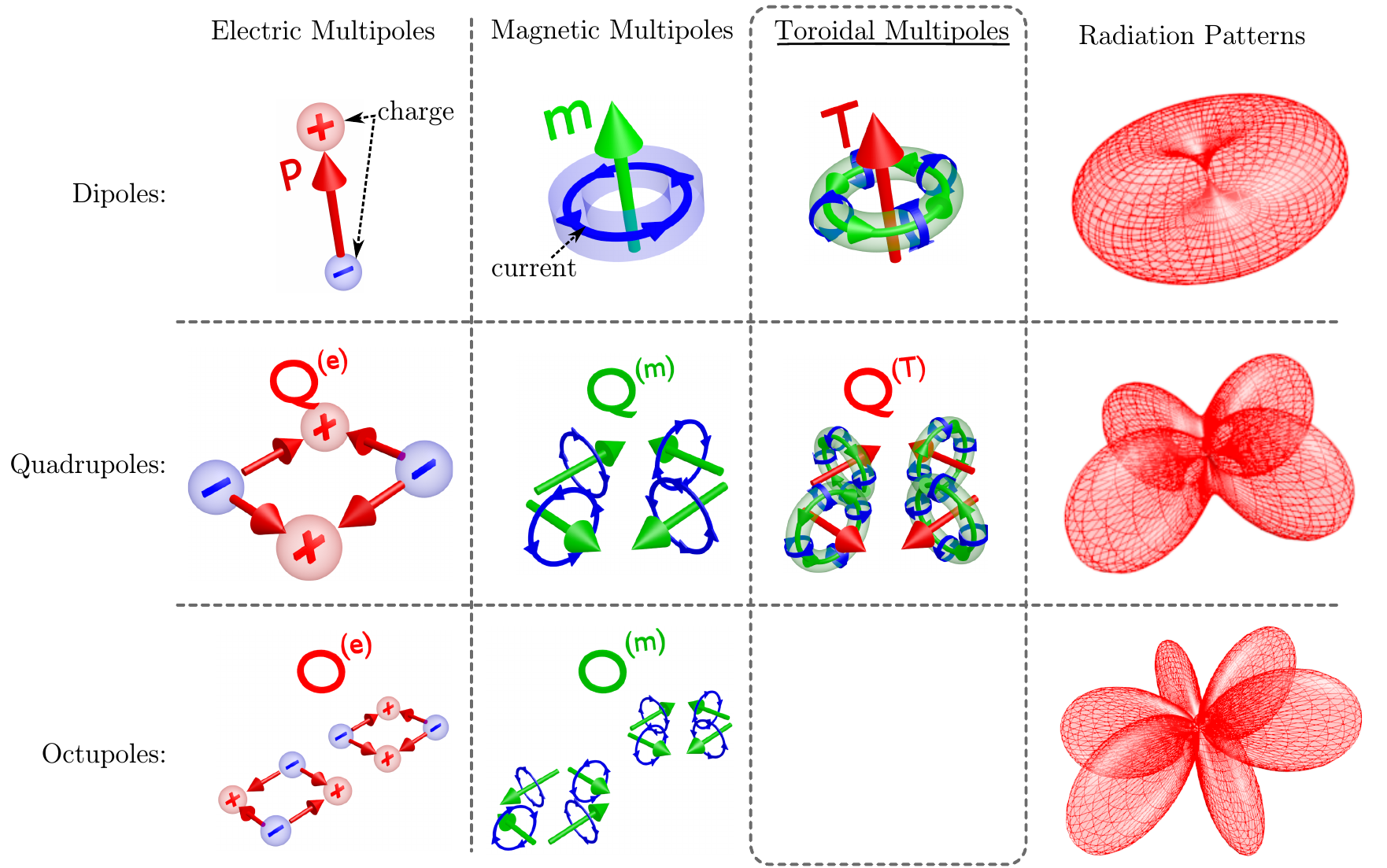}

\caption{\textbf{The members from the three families of dynamic multipoles
up to third order.} The three columns on the left show the charge-current
distribution that give rise to the leading multipoles up to the third
order. The included multipoles are the electric (\textbf{p}), magnetic
(\textbf{m}) and toroidal (\textbf{T}) dipoles, electric ($\mathbf{Q^{(e)}}$),
magnetic ($\mathbf{Q^{(m)}}$) and toroidal ($\mathbf{Q^{(T)}}$)
quadrupoles as well as the electric ($\mathbf{O^{(e)}}$) and magnetic
($\mathbf{O^{(m)}}$) octupoles. The toroidal dipole (\textbf{T}),
is created by the oscillating poloidal current - the current that
flows along the meridians of a torus. The next member of the toroidal
multipole family, the toroidal quadrupole ($\mathbf{Q^{(T)}}$), is
created by the anti-aligned pairs of toroidal dipoles. The column
on the right shows the patterns of radiation (i.e. intensity as a
function of direction) emitted by the various harmonically oscillating
multipoles (for quadrupoles, only the radiation associated with the
off-diagonal component of the second-rank quadrupole moment tensor
is shown, for octupoles only the radiation associated with the component
of the third-rank octupole moment tensor with two repeated indices
and the third distinct index is shown).}

\label{fig:MultipoleZoo}
\end{figure}

\twocolumngrid

The discovery of the toroidal dipole can be traced back to Zel'dovich's
original work\cite{Zeldovich57} in 1957. It is the first member of
the toroidal multipole family and is created by the currents flowing
along the meridians of a torus (see Fig.~\ref{fig:MultipoleZoo}).
Since the discovery, the toroidal dipoles have been used to test the
Standard Model \cite{Wood97}, demonstrate strong optical activity
\cite{Nikitas09}, and negative index of refraction \cite{Marinov07}.
The combination of dynamic magnetic and static toroidal dipoles has
been predicted to violate Newtons Third Law \cite{Afanasiev01}, whilst
different opinions exist on whether the combination of dynamic electric
and toroidal dipoles could lead to a non-radiating configuration,
which emits no electromagnetic radiation despite being a source of
propagating, non-trivial (un-removable by any gauge choice) vector
potential \cite{Afa95,Ziolkowski02,Broadman05,VAFNonRad12}. Following
the recent experimental demonstration of the toroidal dipole response
in metamaterials \cite{Kaelberer10}, the interest in this topic has
been growing rapidly \cite{Dong12,Dong_OptMM_12,Ogut12,Soukoulis12,VAFNonRad12,HuangToro12,Huang13}
despite the lack of theory linking the microscopic toroidal electrodynamics
to macroscopic observables such as transmission and reflection. In
this paper we develop a fully analytical formalism to fill this gap. 

The electromagnetic properties of media are generally described in
terms of macroscopic material parameters (such as, for example, dielectric
permittivity $\epsilon$ and magnetic permeability $\mu$) that through
constitutive relations establish a connection between the media\textquoteright{}s
macroscopic response and microscopic charge-current excitations induced
by the electromagnetic fields in media's constituents, i.e. atoms
or molecules \cite{jackson}. Such description is being also applied
to the so-called metamaterials, man-made material composites with
exotic electromagnetic properties achieved through structuring on
the sub-wavelength scale \cite{NizSci10,Soukolis11,Liu11}. However,
obtaining effective material parameters for the metamaterials is not
straightforward and often impossible due to their structural inhomogeneity
and strong spatial dispersion \cite{Simovski11,Chipouline12}.

Here we present a formalism that allows one to calculate the transmission
and reflection of two-dimensional metamaterials (as well as the films
of sub-wavelength thickness made from conventional materials) based
directly on the multipolar decomposition of the microscopic charge-current
excitations, thus avoiding the need for introducing the effective
material parameters. Similar problem of calculating the scattered
radiation from arrays of metallic resonators have been addressed in
the past using Fast Multipole Method (FMM) \cite{Engheta92,Coifman93,Lu93,Craeye06,Lu07},
and periodic Green\textquoteright{}s functions for the Helmholtz equation
\cite{Shubair93,Fructos04,Valerio07}. What makes our approach different,
is that it yields expressions sufficiently compact to be suitable
not only for computer-aided calculations (like FMM), but also for
the purely analytic evaluation. At the same time our approach accounts
not only for the conventional multipoles but also for the elusive
toroidal multipoles (see Fig.~\ref{fig:MultipoleZoo}). By applying
the derived formalism to a test case study metamaterial, we show that
characterization of the electromagnetic response of a certain class
of structures is greatly enhanced by taking the toroidal multipoles
into account.

We will now proceed to deriving a general expression for the electromagnetic
field scattered by a two-dimensional array of identical charge-current
excitations that are represented by a finite series of dynamic multipoles.
For the case of a passive materials (and metamaterials), these multipoles
would be induced by normally incident plane wave. We assume that the
multipole moments can either be extracted from the numerical simulation
of the induced currents, or can be calculated from the anticipated
dynamics of charge and current densities induced in the meta-molecules
of a particular geometry by the incident radiation \cite{Saadoun94,Bahr94,Pendry99,Rockstuhl07,Sten11}.
In the interest of brevity only the key steps of the derivation will
be demonstrated by finding the expression for the radiation from a
two-dimensional sub-wavelength array of toroidal dipoles, before giving
the full expression that includes all lower-order multipoles. 

We start from the far-field distribution of the electric field radiated
by a single oscillating toroidal dipole, which has been derived by
Radescu \& Vaman in \cite{radescu02} (also see Appendix):

\begin{eqnarray}
\mathbf{E}(\mathbf{r}) & = & \frac{-\II\mu_{0}c^{2}k^{3}}{3\sqrt{2\pi}}\cdot\frac{\exp\left(-\II kr\right)}{r}\cdot\nonumber \\
 &  & \cdot\sum_{m=0,\pm1}T_{1m}\left[\mathbf{Y}_{1,2,m}+\sqrt{2}\mathbf{Y}_{1,0,m}\right]\label{eq:EFieldSingleTDip}\\
T{}_{1,\pm1} & = & \frac{1}{\sqrt{2}}\left(\mp T_{x}+\II T_{y}\right)\label{eq:TDipSpherCartXY}\\
T_{1,0} & = & T_{z}\label{eq:TDipSpherCartZ}\\
\mathbf{T} & = & \frac{1}{10c}\int d^{3}r\,\left[\mathbf{r}\left(\mathbf{r}.\mathbf{J}\right)-2r^{2}\mathbf{J}\right]\label{eq:TDipCartInt}
\end{eqnarray}

Here $\mu_{0}$ is the magnetic permeability of vacuum, $c$ is the
speed of light, $\mathbf{r}$ is the vector connecting the location
of the dipole with the observer and $\mathbf{Y}_{l,l',m}$ are the
spherical vector harmonics that allow to represent any vector field
on the surface of the unit sphere in the same way as spherical harmonics
allow to represent any scalar field on the surface of the unit sphere
\cite{QEDBook,Dubovik75,radescu02}. The (Cartesian) toroidal dipole
is denoted by $\mathbf{T}$, whilst $\mathbf{J}$ is the current density
that gives rise to dipole. Unlike Vaman \& Radescu \cite{radescu02},
we are using the SI units and assume harmonic time-dependence specified
by $\sim\exp(+\II\omega t)$, where $\omega$ is the angular frequency
and $k=2\pi/\lambda=\omega/c$ is the wavenumber.

The total field radiated by an infinitely large planar array of toroidal
dipoles ($\mathbf{E}_{s}$) is obtained by summing the contributions
from all the dipoles at the position of the observer. As stated above,
we assume that all the dipoles oscillate in phase (i.e. the multipole
array is induced by the plane wave at normal incidence), and that
the unit cell of the array is smaller than the wavelength. The latter
assumption allows to replace the sum over the unit cells with an integral
over the array area ($\Delta^{2}$ denotes the area of the unit cell).
\begin{equation}
\mathbf{E_{s}}=\sum_{\mathbf{r}}\mathbf{E}(\mathbf{r})\approx\frac{1}{\Delta^{2}}\int d^{2}r\,\mathbf{E}(\mathbf{r})\label{eq:TDipPlaneWave}
\end{equation}

\begin{figure}
\includegraphics{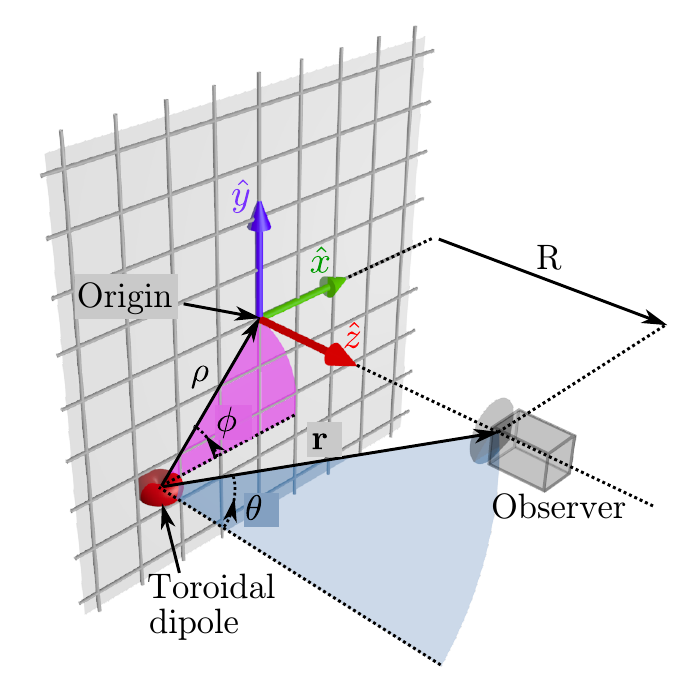}

\caption{\textbf{Calculating scattering from a two-dimensional planar array
of toroidal dipoles.} A single toroidal dipole is represented by its
radiation pattern. The vectors connecting the dipole to the observer
and to the origin of the array are $\mathbf{r}$ and $\mathbf{\rho}$,
respectively. The observer is located at distance $R$ from the array.
The vector $\mathbf{R}$ is perpendicular to the array plane. The
position of the observer relative to the dipole in spherical polar
coordinates is $(r,\,\theta,\,\phi)$. }

\label{fig:TDipArrayDiag}
\end{figure}

We choose to work in the coordinate system where the array of dipoles
lies in the xy-plane and the incident/scattered radiation propagates
along the z-axis (see Fig.~\ref{fig:TDipArrayDiag}). Explicit evaluation
of the relevant spherical vector harmonics produces \cite{radescu02}:
\begin{eqnarray*}
\mathbf{Y}_{1,2,\pm1}+\sqrt{2}\mathbf{Y}_{1,0,\pm1}=\\
=\left(\begin{array}{c}
\pm\sqrt{\frac{3}{10}}Y_{2,\pm2}\mp\sqrt{\frac{1}{20}}Y_{2,0}\mp Y_{0,0}\\
-\II\sqrt{\frac{3}{10}}Y_{2,\pm2}-\II\sqrt{\frac{1}{20}}Y_{2,0}-\II Y_{0,0}\\
-\sqrt{\frac{3}{10}}Y_{2,\pm1}
\end{array}\right)
\end{eqnarray*}
\[
\mathbf{Y}_{1,2,0}+\sqrt{2}\mathbf{Y}_{1,0,0}=\left(\begin{array}{c}
\sqrt{\frac{3}{20}}Y_{2,1}-\sqrt{\frac{3}{20}}Y_{2,-1}\\
-\II\sqrt{\frac{3}{20}}Y_{2,1}-\II\sqrt{\frac{3}{20}}Y_{2,-1}\\
-\sqrt{\frac{2}{5}}Y_{2,0}+\sqrt{2}Y_{0,0}
\end{array}\right)
\]

The vectors are presented in the Cartesian basis with column entries
indicating the x-, y- and z-components (from top to bottom respectively),
$Y_{l,m}$ are the standard spherical harmonics \cite{radescu02}.
The basic integral that needs to be calculated in Eq.~(\ref{eq:TDipPlaneWave}),
after substitution of Eq.~(\ref{eq:EFieldSingleTDip}), is: 
\begin{equation}
I_{l,m}=\int d^{2}r\, Y_{l,m}\exp\left(-\II kr\right)/r\label{eq:CoreSphericalHarmInt}
\end{equation}

By assuming that the propagation of radiation occurs in space with
losses, i.e. $\Im(k)<0$, and by focusing on the far-field component
of radiation, i.e. by assuming that the distance between the observer
and the dipole array is significantly larger than the wavelength of
radiation $R\gg\lambda$, one can show that (see Appendix):
\begin{equation}
I_{l,m}\approxeq\frac{\pi\delta_{m,0}\left(\mathbf{\hat{R}}.\mathbf{\hat{z}}\right)^{l}}{\II k}\cdot\sqrt{\frac{2l+1}{\pi}}\cdot\exp\left(-\II kR\right)\label{eq:CoreSphericalHarmInt_Solved}
\end{equation}

where $\mathbf{\hat{R}}=\mathbf{R}/R$ (see Fig.~\ref{fig:TDipArrayDiag}).
Substitution of Eq.~(\ref{eq:EFieldSingleTDip}) into Eq.~(\ref{eq:TDipPlaneWave})
and use of Eq.~(\ref{eq:CoreSphericalHarmInt_Solved}) allows to
derive:
\[
\mathbf{E_{s}}=\frac{\mu_{0}c^{2}k^{2}}{4\Delta^{2}}\cdot\sqrt{2}\cdot\left(\begin{array}{c}
T_{1,1}-T_{1,-1}\\
\II\left(T_{1,1}+T_{1,-1}\right)\\
0
\end{array}\right)\cdot\exp\left(-\II kR\right)
\]

Further simplification produces the final form:
\begin{equation}
\mathbf{E_{s}}=-\frac{\mu_{0}c^{2}k^{2}}{2\Delta^{2}}\,\mathbf{T}_{\parallel}\,\exp\left(-\II kR\right)\label{eq:EFieldMDipFinal}
\end{equation}

where $\mathbf{T}_{\parallel}=\mathbf{T}-(\mathbf{T}.\mathbf{\hat{R}})\mathbf{\hat{R}}$
denotes the projection of toroidal dipole into the plane of the array
($\mathbf{T}_{\parallel}=\left(T_{x},T_{y},0\right)^{T}$ for the
coordinate system as in Fig.~\ref{fig:TDipArrayDiag}). Repetition
of the derivation given above for other multipoles results in:

\begin{eqnarray}
\mathbf{E}_{s}=\frac{\mu_{0}c^{2}}{2\Delta^{2}}\cdot\left[-\II k\mathbf{p}_{\parallel}+\II k\mathbf{\hat{R}}\times\left(\mathbf{m}_{\parallel}-\frac{k^{2}}{10}\mathbf{m}_{\parallel}^{\left(1\right)}\right)-\right.\nonumber \\
-k^{2}\left(\mathbf{T}_{\parallel}+\frac{k^{2}}{10}\mathbf{T}_{\parallel}^{\left(1\right)}\right)+k^{2}\left(\mathbf{Q^{(e)}}.\mathbf{\hat{R}}\right)_{\parallel}-\nonumber \\
-\frac{k^{2}}{2}\mathbf{\hat{R}}\times\left(\mathbf{Q^{(m)}}.\mathbf{\hat{R}}\right)_{\parallel}-\frac{\II k^{3}}{3}\left(\mathbf{Q^{(T)}}.\mathbf{\hat{R}}\right)_{\parallel}+\nonumber \\
+\II k^{3}\left(\left(\mathbf{O^{(e)}}.\mathbf{\hat{R}}\right).\mathbf{\hat{R}}\right)_{\parallel}-\nonumber \\
\left.-\frac{\II k^{3}}{180}\mathbf{\hat{R}}\times\left(\left(\mathbf{O^{(m)}}.\mathbf{\hat{R}}\right).\mathbf{\hat{R}}\right)_{\parallel}\right]\cdot\nonumber \\
\cdot\exp\left(-\II kR\right)\quad\label{eq:FullMultField}
\end{eqnarray}

Equation~(\ref{eq:FullMultField}) describes electric field emitted
by an infinitely large two-dimensional array of meta-molecules (or
any sub-wavelength emitters) with induced oscillations of charge-current
density approximated by first 8 dynamic multipoles. It contains 10
terms corresponding to the electric ($\mathbf{p}$), toroidal ($\mathbf{T}$)
and magnetic ($\mathbf{m}$) dipoles, electric ($\mathbf{Q^{(e)}}$),
magnetic ($\mathbf{Q^{(m)}}$) and toroidal ($\mathbf{Q^{(T)}}$)
quadrupoles, electric ($\mathbf{O^{(e)}}$) and magnetic ($\mathbf{O^{(m)}}$)
octupoles, and the so-called mean square radii of toroidal ($\mathbf{T}^{(1)}$)
and magnetic ($\mathbf{m}^{(1)}$) dipoles, which are the lowest-order
corrections retained to account for the finite size of the meta-molecules
\cite{radescu02}. Further multipole contributions to the radiation
by an array can be found in the same way. 

Using Eq.~(\ref{eq:FullMultField}), the radiation transmitted and
reflected by the two-dimensional array, when it is subjected to illumination
by normally incident plane wave, can be found from: 

\begin{eqnarray*}
\mathbf{E}_{reflected} & = & \left[\mathbf{E}_{s}\right]{}_{\mathbf{\hat{R}}=-\mathbf{\hat{k}}}\\
\mathbf{E}_{transmitted} & = & \left[\mathbf{E}_{s}\right]{}_{\mathbf{\hat{R}}=\mathbf{\hat{k}}}+\mathbf{E}_{incident}
\end{eqnarray*}

where $\mathbf{\hat{k}}=\mathbf{k}/k$ points in the direction of
propagation of the incident radiation. 

\begin{figure}
\includegraphics{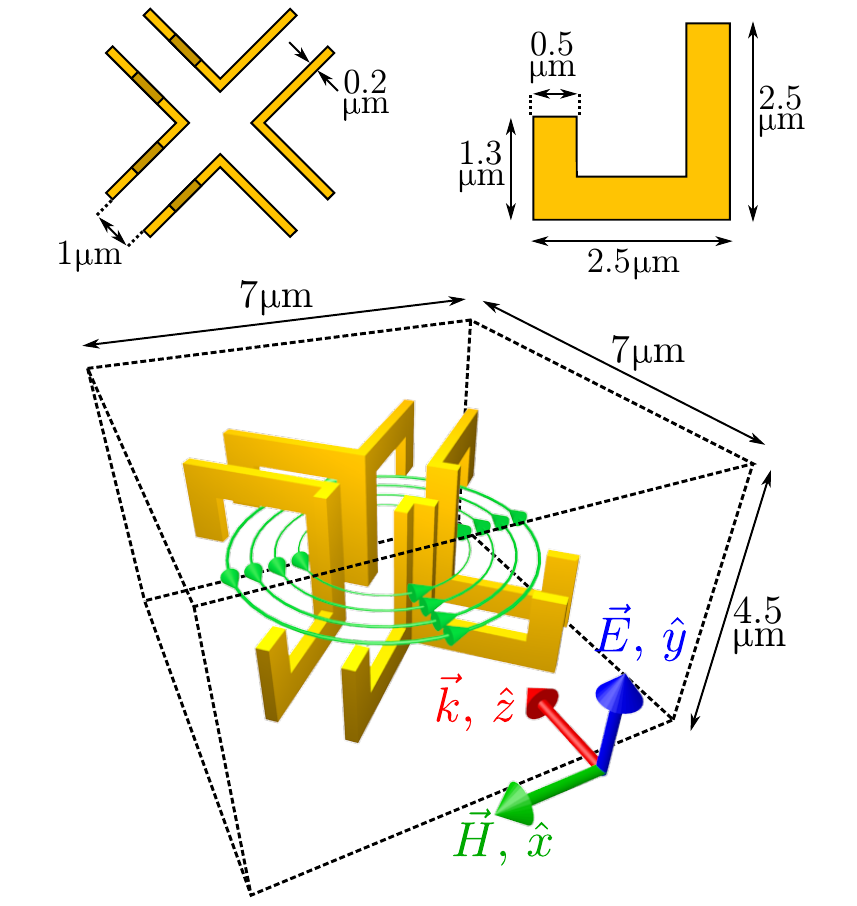}

\caption{\textbf{The unit cell of the test case toroidal metamaterial. }The
metamaterial is created by translating the unit cell along the $\mathbf{\hat{x}}$
and $\mathbf{\hat{y}}$ directions. The unit cell consists of four
pairs of gold split-ring resonators that are embedded in the SU-8
polymer. The metamaterial is driven by the radiation propagating along
the $\mathbf{\hat{z}}$-axis and polarized along $\mathbf{\hat{y}}$-axis.
The configuration of the magnetic field that gives rise to the toroidal
dipole response in the metamaterial is schematically illustrated with
green field-lines. }

\label{fig:MMDiag}
\end{figure}

Below we will illustrate application of our approach for calculation
of electromagnetic response of a metamaterial designed to exhibit
strong toroidal resonance in the mid-IR part of the spectrum. The
unit cell of the metamaterial array, shown in Fig.~\ref{fig:MMDiag},
contains a three-dimensional complex-shaped gold meta-molecule with
the main features of the four-split-ring design proposed by Kaelberer
\emph{et al.} \cite{Kaelberer10}. The current design was optimized
for the novel metamaterial fabrication technique SAMPL \cite{sandia10},
resulting in each of the four split rings being replaced with a pair
of split rings of highly asymmetric shape, to reduce the effect of
losses and maximize the contribution of the toroidal dipole moment.
The transmission and reflection of the array of such meta-molecules
were simulated in $14.5\,\upmu\mbox{m}-23.5\,\upmu\mbox{m}$ wavelength
range (see Appendix for material constants) using full 3D Maxwell\textquoteright{}s
equations solver (COMSOL Multiphysics 3.5a). The numerical model also
provided data on spatial distribution of the current densities, which
were used to calculate dynamic multipole moments induced in each meta-molecule
(see Appendix).

\begin{figure}
\includegraphics{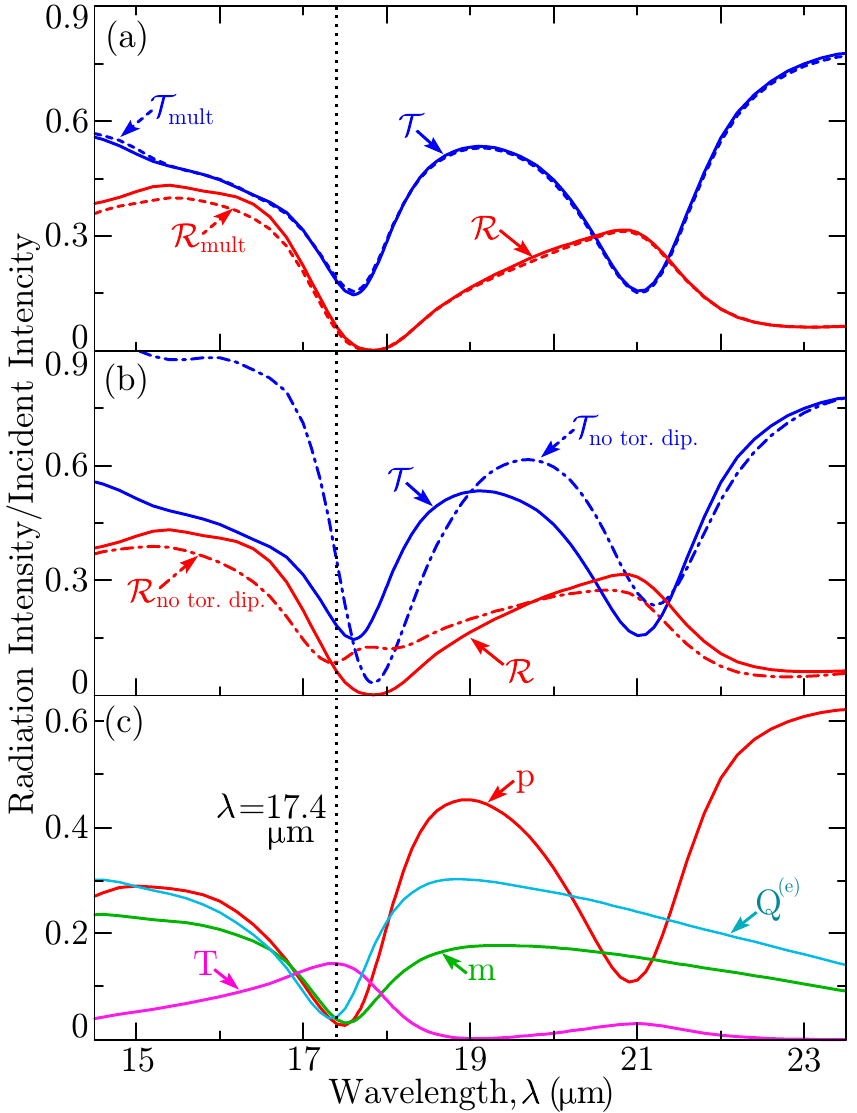}

\caption{\textbf{Macroscopic response of the test case toroidal metamaterial}.
The response of the metamaterial on Fig.~\ref{fig:MMDiag} has been
modeled using the realistic material constants for gold and SU-8 polymer
(see Appendix). \textbf{(a)} Transmission ($\mathcal{T}$; blue) and
reflection ($\mathcal{R}$; red) of the metamaterial. The solid curves
($\mathcal{T}$, $\mathcal{R}$) are obtained directly from numerical
solution of Maxwell's equations, dashed curves ($\mbox{\ensuremath{\mathcal{T}}}_{\mathrm{{mult}}},\,\mbox{\ensuremath{\mathcal{R}}}_{\mathrm{{mult}}}$)
are obtained using Eq.~(\ref{eq:FullMultField}), from the multipole
moment representation of each meta-molecule of the metamaterial. \textbf{(b)
}The comparison of the directly calculated metamaterial transmission
and reflection ($\mathcal{T}$,~$\mathcal{R}$; solid curves) to
the response obtained from multipoles but without the input of toroidal
dipole ($\mathcal{T}_{\mathrm{{no\, tor.\, dip.}}},\,\mathcal{R}_{\mathrm{{no\, tor.\, dip.}}}$;
dash-dot curves). \textbf{(c)} Intensity of the radiation scattered
back (i.e. reflected) by each of four leading multipoles when placed
in an array (\textbf{p}-electric dipole, \textbf{m}-magnetic dipole,
\textbf{T}-toroidal dipole, \textbf{$\mathbf{Q^{(e)}}$}-electric
quadrupole). The radiation intensity in all three plots is normalized
with respect to incident radiation (that induces the multipoles).}

\label{fig:MM_TR}
\end{figure}

The simulated transmission and reflection spectra are shown in Fig.~\ref{fig:MM_TR}a,b
as solid curves, revealing two distinct resonances located at around
$21.0\,\upmu\mbox{m}$ and $17.4\,\upmu\mbox{m}$. The numerical spectra
are very well matched by the results of the multipole calculations
described above (see dashed curves in Fig.~\ref{fig:MM_TR}a). Small
discrepancies are attributed to somewhat limited accuracy of extracting
the induced current distribution from the numerical model. The shorter-wavelength
resonance corresponds to strong toroidal response, which is confirmed
by our analysis of the multipole scattering presented in Fig.~\ref{fig:MM_TR}c
(only four leading multipoles are shown). It shows that, for each
meta-molecule, the power scattered by the induced toroidal dipole
at $17.4\,\upmu\mbox{m}$ is more than three times larger than the
contribution from any standard multipole, and therefore toroidal dipole
excitation must play the key role in forming the metamaterial macroscopic
response at this wavelength. This can be verified directly by excluding
the toroidal dipole moment from the multipole-based calculations of
the transmission and reflection. As one can see from Fig.~\ref{fig:MM_TR}b
the correct replication of the resonant features is simply not possible
in the frame of the standard multipole expansion, and the notion of
the toroidal dipole is thus crucial for the correct interpretation
of the macroscopic response of metamaterial.

In conclusion, we developed a fully analytical formalism that allows
calculating the transmission and reflection properties of thin sheets
of metamaterials and material composites, based on the dynamic multipole
decomposition of charge-current densities induced in their structure
by an incident electromagnetic wave. Further to the derived formalism,
we provided a case study which proved that the contribution of the
toroidal dipole is crucial for the correct interpretation of the reflection
and the transmission spectra of a certain class of metamaterials.
Our findings demonstrate that the toroidal dipole may be dominant
in the response of the electromagnetic media, and therefore cannot
be treated simply as a high-order correction to the electric or magnetic
multipoles.
\begin{acknowledgments}
The authors acknowledge the support of the Engineering and Physical
Sciences Research Council U.K., the Royal Society and of the MOE Singapore
grant MOE2011-T3-1-005. The authors would also like to thank Dr. I.
Brener and Dr. D. B. Burckel from Sandia National Laboratories (US)
for providing the material data and discussions. 
\end{acknowledgments}

\section*{Appendix}

\subsection*{Integral involving the Spherical Harmonics - $I_{l,m}$}

Here we will derive the Eq.~(7). At the core of the derivation lies
the evaluation of the following integral:
\begin{equation}
\int_{R}^{\infty}dr\,\left(\frac{R}{r}\right)^{q}\exp\left(-\II kr\right)\approxeq\frac{\exp\left(-\II kR\right)}{\II k},\quad\Im\left(k\right)<0\label{eq:App:CoreInt}
\end{equation}

The case $q=0$ can be found by the direct integration. Higher order
cases can be evaluated by relating them to the exponential integrals.
Abramowitz and Stegun\cite{Abramowitz} define the exponential integral
as (Eq.~(5.1.4) of Ref.~{[}\onlinecite{Abramowitz}{]}):
\[
E_{n}(z)=\int_{1}^{\infty}dt\,\frac{\exp\left(-zt\right)}{t^{n}},\quad n=0,\,1,\,2,...\quad\Re\left(z\right)>0
\]

We are interested in the asymptotic expansion of the $E_{n}(z)$ for
the case of large $z$ given in Eq.~(5.1.51) of Ref.~{[}\onlinecite{Abramowitz}{]}:
\[
\lim_{z\to\infty}E_{n}(z)\approxeq\frac{\exp\left(-z\right)}{z}\left(1-O\left(1/z\right)\right),\quad\left|\arg\left(z\right)\right|<\frac{3}{2}\pi
\]

Equation~ (\ref{eq:App:CoreInt}) can therefore as be evaluated as
follows:

\begin{eqnarray*}
\int_{R}^{\infty}dr\,\left(\frac{R}{r}\right)^{q}\exp\left(-\II kr\right)=R\cdot E_{q}(\II kR)\approxeq\\
\approxeq R\cdot\frac{\exp\left(-\II kR\right)}{\II kR}\cdot\left(1-O\left(1/kR\right)\right)
\end{eqnarray*}

Note that $\Im(k)<0$ implies $\Re(\II kR)>0$ and $\left|\arg(\II kR)\right|<\pi/2$.
Up to order $O\left(1/kR\right)$ or, equivalently, up to $O\left(\lambda/R\right)$,
the expression becomes:
\[
\int_{R}^{\infty}dr\,\left(\frac{R}{r}\right)^{q}\exp\left(-\II kr\right)\approxeq\frac{\exp\left(-\II kR\right)}{\II k}
\]

We now turn our attention to Eq.~(\ref{eq:CoreSphericalHarmInt}):
\[
I_{l,m}=\int d^{2}r\, Y_{l,m}(\theta,\phi)\exp\left(-\II kr\right)/r
\]

\begin{figure}
\includegraphics{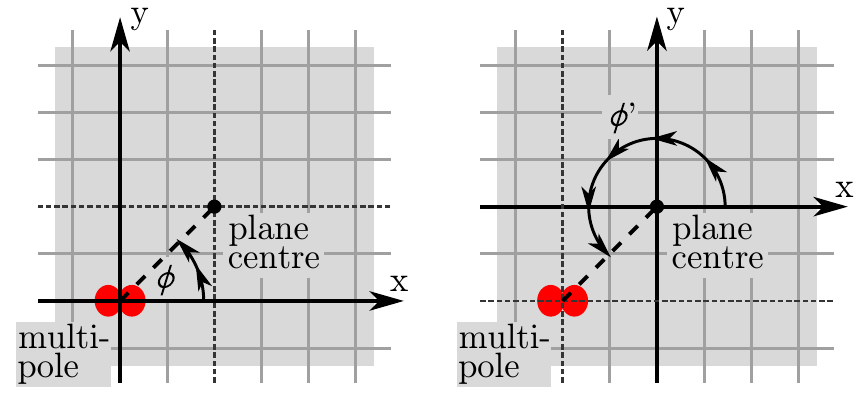}

\caption{\textbf{The difference between $\mathbf{\phi}$ and $\mathbf{\phi'}$.}
Angle $\phi$ is defined in the coordinate frame of a multipole under
consideration, whilst the angle $\phi'$ is defined in the coordinate
frame of the multipole array (also see Fig.~2). The observer (see
Fig.~2) is located directly above the origin of the array.}

\label{fig:PhiAndPhiDash}
\end{figure}

The integration is understood to be over the area of the array of
multipoles as shown in Fig.~\ref{fig:TDipArrayDiag}. The position
of each multipole in the plane of the array is given by $\rho$, the
distance between the centre-point of the array and the considered
multipole, and $\phi$' the angle between the x-axis and the vector
connecting the centre-point of the array and the multipole in question.
There is also another angle $\phi$ that belongs together with $r$
and $\theta$, and denotes the position of the observer relative to
the multipole under consideration (see Fig.~\ref{fig:TDipArrayDiag}).
It is convenient to place the origin of the multipole array directly
below the observer. In this case the the relation between $\phi$
and $\phi'$ takes a simple form $\phi=\phi'+\pi$, up to a full rotation
around $2\pi$. Figure~\ref{fig:PhiAndPhiDash} helps to visualize
the two angles. The same choice of origin establishes the relation
$r^{2}=\rho^{2}+R^{2}$. 

One can now rewrite the integral in more accessible way:
\[
I_{l,m}=\int_{0}^{2\pi}d\phi'\,\int_{0}^{\infty}\rho d\rho\, Y_{l,m}\left(\theta,\phi'+\pi\right)\frac{\exp\left(-\II kr\right)}{r}
\]

From $r^{2}=\rho^{2}+R^{2}$ it follows that $r\, dr=\rho\, d\rho$,
so 
\begin{eqnarray*}
I_{l,m} & = & \int_{0}^{2\pi}d\phi'\,\int_{R}^{\infty}rdr\, Y_{l,m}\left(\theta,\phi'+\pi\right)\frac{\exp\left(-\II kr\right)}{r}\\
 & = & \left(-1\right)^{m+m}\sqrt{\frac{2l+1}{4\pi}\,\frac{(l-m)!}{(l+m)!}}\cdot\\
 &  & \cdot\int d\phi'\exp\left(\II m\phi'\right)\int_{R}^{\infty}dr\, P_{l}^{m}\left(\cos\theta\right)\exp\left(-\II kr\right)
\end{eqnarray*}

In the last step, we have expanded the spherical harmonic following
the convention used by Arfken and Weber (see Chapter~12.6 in Ref.~{[}\onlinecite{ArfkenWeber}{]}),
and substituted $\exp\left(\II m\left(\pi+\phi'\right)\right)=\left(-1\right)^{m}\exp\left(\II m\phi'\right)$.
Here the $P_{l}^{m}$ denotes the Associated Legendre Functions. The
expression above is simplified considerably by the fact that the integral
over $\phi'$ is non-zero only for $m=0$:
\[
I_{l,m}=\pi\delta_{m,0}\sqrt{\frac{2l+1}{\pi}}\,\int_{R}^{\infty}dr\, P_{l}\left(\cos\theta\right)\exp\left(-\II kr\right)
\]

Above we have used $P_{l}^{0}(x)=P_{l}(x)$ to replace the Associated
Legendre Functions with Legendre Polynomials (respectively). From
Fig.~\ref{fig:TDipArrayDiag} it follows that $\cos\theta=R/r$ for
$\mathbf{R}=R\mathbf{\hat{z}}$, and $\cos\theta=-R/r$ for $\mathbf{R}=-R\mathbf{\hat{z}}$,
thus $\cos\theta=\left(\mathbf{\hat{R}}.\mathbf{\hat{z}}\right)\times R/r$.
Using the parity property of Legendre Polynomials (Eq.~(12.37) in
Ref.~{[}\onlinecite{ArfkenWeber}{]}) one obtains $P_{l}\left(\cos\theta\right)=\left(\mathbf{\hat{R}}.\mathbf{\hat{z}}\right)^{l}\, P_{l}\left(R/r\right)$.
Being a polynomial $P_{l}(x)$ can be expressed as power series $P_{l}(x)=\sum_{s=0}^{\infty}a_{s}^{(l)}x^{s}$,
the integral then becomes (with use of Eq.~(\ref{eq:App:CoreInt})):

\begin{eqnarray*}
I_{l,m} & = & \pi\delta_{m,0}\sqrt{\frac{2l+1}{\pi}}\,\left(\mathbf{\hat{R}}.\mathbf{\hat{z}}\right)^{l}\cdot\\
 &  & \cdot\sum_{s=0}^{\infty}a_{s}^{(l)}\int_{R}^{\infty}dr\,\left(\frac{R}{r}\right)^{s}\exp\left(-\II kr\right)\approxeq\\
 & \approxeq & \pi\delta_{m,0}\sqrt{\frac{2l+1}{\pi}}\,\left(\mathbf{\hat{R}}.\mathbf{\hat{z}}\right)^{l}\sum_{s=0}^{\infty}a_{s}^{(l)}\cdot\\
 &  & \cdot\left(\frac{\exp\left(-\II kR\right)}{\II k}+O\left(\lambda/R\right)\right)
\end{eqnarray*}

Finally, one uses the normalization of the Legendre Polynomials to
eliminate the sum $P_{l}(1)=1=\sum_{s=0}^{\infty}a_{s}^{(l)}$ (Eq.~(12.31)
in Ref.~{[}\onlinecite{ArfkenWeber}{]}). Thus 
\[
I_{l,m}\approxeq\frac{\pi\delta_{m,0}\left(\mathbf{\hat{R}}.\mathbf{\hat{z}}\right)^{l}}{\II k}\cdot\sqrt{\frac{2l+1}{\pi}}\cdot\exp\left(-\II kR\right)
\]

which completes the derivation.

\subsection*{Multipole decomposition of the radiation from a localized source}

To derive the formula for the electric field radiated by the array
of multipoles (see Eq.~(\ref{eq:FullMultField})) we have used the
expression for the radiation emitted by the single multipole sources
provided by Radescu and Vaman (see Eq.~(3.15) in Ref.~{[}\onlinecite{radescu02}{]}).
Here we will give the truncated series for the electric field emitted
by the multipole sources, in the SI units, and for the complex-valued
harmonic time-dependence of the source ($\sim\exp\left(+\II\omega t\right)$).

Due to large number of terms it is convenient to separate the series
into different orders of $l$. The $l=1$ sub-series then contain
the dipolar contributions:

\begin{eqnarray*}
\mathbf{E}_{(l=1)} & \approx & \frac{\mu_{0}c^{2}}{3\sqrt{2\pi}}\cdot\frac{\exp\left(-ikr\right)}{r}\cdot\sum_{m=0,\pm1}\Big[\\
 &  & \left(k^{2}Q_{1,m}-\II k^{3}T_{1,m}+\II k^{5}T_{1,m}^{(1)}\right)\cdot\\
 &  & \cdot\left(\mathbf{Y}_{1,2,m}+\sqrt{2}\mathbf{Y}_{1,0,m}\right)+\\
 &  & +\II\sqrt{3}\left(k^{2}M_{1.m}-k^{4}M_{1,m}^{(1)}\right)\cdot\mathbf{Y}_{1,1,m}\Big]
\end{eqnarray*}

$l=2$ sub-series contain the quadrupolar contributions:

\begin{eqnarray*}
\mathbf{E}_{(l=2)} & \approx & \frac{\mu_{0}c^{2}}{10\sqrt{6\pi}}\cdot\frac{\exp\left(-ikr\right)}{r}\cdot\sum_{m=0,\pm1,\pm2}\Big[\\
 &  & \left(\II k^{3}Q_{2,m}^{(e)}+k^{4}Q_{2,m}^{(T)}\right)\cdot\\
 &  & \left(\sqrt{2}\mathbf{Y}_{2,3,m}+\sqrt{3}\mathbf{Y}_{2,1,m}\right)-\\
 &  & -\sqrt{5}k^{3}Q_{2,m}^{(m)}\mathbf{Y}_{2,2,m}\Big]
\end{eqnarray*}

$l=3$ sub-series contain the octupolar contributions:
\begin{eqnarray*}
\mathbf{E}_{(l=3)} & \approx- & \frac{\mu_{0}c^{2}k^{4}}{15\sqrt{3\pi}}\cdot\frac{\exp\left(-ikr\right)}{r}\cdot\sum_{m=0,\pm1,\pm2,\pm3}\Big[\\
 &  & \frac{1}{7}O_{3,m}^{(e)}\cdot\left(\sqrt{3}\mathbf{Y}_{3,4,m}+2\mathbf{Y}_{3,2,m}\right)+\\
 &  & +\frac{\II}{\sqrt{7}}O_{3,m}^{(m)}\mathbf{Y}_{3,3,m}\Big]
\end{eqnarray*}

The total field emitted is given by:
\[
\mathbf{E}=\mathbf{E}_{(l=1)}+\mathbf{E}_{(l=2)}+\mathbf{E}_{(l=3)}+(l>3\mbox{\,\ sub-series})
\]

The series given above are truncated at order $k^{4}$, but the first-order
correction for the toroidal dipole ($T_{1,m}^{(1)}$), of order $k^{5}$,
is also included to avoid errors in the spectral range where toroidal
dipole dominates (see Fig.~\ref{fig:MM_TR}). The other $k^{5}$
terms that can be included are the toroidal octupole, the electric
and magnetic hexadecapoles ($l=4$), and the first-order correction
to the magnetic quadrupole. 

One may notice, that no correction terms for the electric dipoles
have been included. Radescu and Vaman\cite{radescu02} have shown
that the correction terms for the electric multipoles do not contribute
to the far-field radiation emitted by arbitrary localized charge-current
density distributions. The correction terms for the magnetic and toroidal
multipoles, by contrast, do contribute.

\subsection*{Integrals for finding the leading multipoles from a current distribution}

The expressions we have used to calculate the multipole moments from
the current density distribution are those given by Radescu and Vaman\cite{radescu02}.
We will repeat them here for convenience. Note that the electric and
magnetic multipoles are exactly the same as the ones given in the
standard texts on electrodynamics\cite{jackson} (apart from the different
normalization constants).

Cartesian multipoles are computed by integrating over the charge density
($\rho(\mathbf{r})$) or current density ($\mathbf{J}(\mathbf{r})$)
distribution within the unit cell ($\alpha,\beta,\gamma=x,y,z$):
\begin{eqnarray*}
p_{\alpha} & = & \int d^{3}r\,\rho\, r_{\alpha}=\frac{1}{\II\omega}\int d^{3}r\, J_{\alpha}\\
m_{\alpha} & = & \frac{1}{2c}\int d^{3}r\,\left[\mathbf{r}\times\mathbf{J}\right]_{\alpha}\\
m_{\alpha}^{(1)} & = & \frac{1}{2c}\int d^{3}r\,\left[\mathbf{r\times J}\right]_{\alpha}r^{2}\\
T_{\alpha} & = & \frac{1}{10c}\int d^{3}r\,\left[\left(\mathbf{r.J}\right)r_{\alpha}-2r^{2}J_{\alpha}\right]\\
T_{\alpha}^{(1)} & = & \frac{1}{28c}\int d^{3}r\,\left[3r^{2}J_{\alpha}-2r_{\alpha}\mathbf{\left(r.J\right)}\right]r^{2}\\
Q_{\alpha,\beta}^{(e)} & = & \frac{1}{2}\int d^{3}r\,\rho\,\left[r_{\alpha}r_{\beta}-\frac{1}{3}\delta_{\alpha,\beta}r^{2}\right]=\\
 & = & \frac{1}{\II2\omega}\int d^{3}r\,\left[r_{\alpha}J_{\beta}+r_{\beta}J_{\alpha}-\frac{2}{3}\delta_{\alpha,\beta}\left(\mathbf{r.J}\right)\right]\\
Q_{\alpha,\beta}^{(m)} & = & \frac{1}{3c}\int d^{3}r\,\left[\mathbf{r\times J}\right]_{\alpha}r_{\beta}+\left\{ \alpha\leftrightarrow\beta\right\} \\
Q_{\alpha,\beta}^{(T)} & = & \frac{1}{28c}\int d^{3}r\,\Big[4r_{\alpha}r_{\beta}\left(\mathbf{r.J}\right)-5r^{2}(r_{\alpha}J_{\beta}+r_{\beta}J_{\alpha})+\\
 &  & +2r^{2}\left(\mathbf{r.J}\right)\delta_{\alpha,\beta}\Big]\\
O_{\alpha,\beta,\gamma}^{(e)} & = & \frac{1}{6}\int d^{3}r\,\rho\, r_{\alpha}\left(r_{\beta}r_{\gamma}-\frac{1}{5}r^{2}\delta_{\beta,\gamma}\right)+\\
 &  & +\left\{ \alpha\leftrightarrow\beta,\gamma\right\} +\left\{ \alpha\leftrightarrow\gamma,\beta\right\} \\
 & = & \frac{1}{\II6\omega}\int d^{3}r\,\bigg[J_{\alpha}\left(r_{\beta}r_{\gamma}-\frac{r^{2}}{5}\delta_{\beta,\gamma}\right)+\\
 &  & +r_{\alpha}\left(J_{\beta}r_{\gamma}+r_{\beta}J_{\gamma}-\frac{2}{5}\mathbf{(r.J)}\delta_{\beta,\gamma}\right)\bigg]+\\
 &  & +\left\{ \alpha\leftrightarrow\beta,\gamma\right\} +\left\{ \alpha\leftrightarrow\gamma,\beta\right\} \\
O_{\alpha,\beta,\gamma}^{(m)} & = & \frac{15}{2c}\int d^{3}r\,\left(r_{\alpha}r_{\beta}-\frac{r^{2}}{5}\delta_{\alpha,\beta}\right)\cdot\left[\mathbf{r\times J}\right]_{\gamma}+\\
 &  & +\left\{ \alpha\leftrightarrow\beta,\gamma\right\} +\left\{ \alpha\leftrightarrow\gamma,\beta\right\} 
\end{eqnarray*}

For quadrupoles and octupoles a short-hand has been used to improve
clarity. For example: $\int d^{3}r\,\left[\mathbf{r\times J}\right]_{\alpha}r_{\beta}+\left\{ \alpha\leftrightarrow\beta\right\} \equiv\int d^{3}r\,\left[\mathbf{r\times J}\right]_{\alpha}r_{\beta}+\int d^{3}r\,\left[\mathbf{r\times J}\right]_{\beta}r_{\alpha}$,
i.e. the second term is obtained from the first term, with the exchanged
positions of indices $\alpha$ and $\beta$. In case of octupoles
(for example):
\[
\frac{1}{6}\int d^{3}r\,\rho\, r_{\alpha}\left(r_{\beta}r_{\gamma}-\frac{1}{5}r^{2}\delta_{\beta,\gamma}\right)+\left\{ \alpha\leftrightarrow\beta,\gamma\right\} +\left\{ \alpha\leftrightarrow\gamma,\beta\right\} 
\]

means that the second term is obtained from the first term by exchanging
$\alpha$ and $\beta$ whilst leaving $\gamma$ untouched. The third
term is, again, obtained from the first term, but this time $\alpha$
and $\gamma$ are exchanged, whilst $\beta$ remains untouched.

In the time-harmonic case there is no clear difference between the
conduction and displacement currents. In simulations we have used
$\mathbf{J}=\II\omega\epsilon_{0}\left(\tilde{\epsilon}_{r}-1\right)\mathbf{E}$
to find the current density within the media, from the electric field
distribution $\mathbf{E}(\mathbf{r})$. The relevant quantities are:
$\omega$-angular frequency, $\epsilon_{0}$-free-space permittivity,
$c$-speed of light, and $\tilde{\epsilon}_{r}$-complex-valued dielectric
constant (used to describe both the dielectrics and metals).

The spherical multipoles are related to the Cartesian multipoles through:
\begin{eqnarray*}
Q_{1,0}=p_{z},\: & Q_{1,\pm1}=\left(\mp p_{x}+\II p_{y}\right)/\sqrt{2}\\
M_{1,0}=-m_{z},\: & M_{1,\pm1}=\left(\pm m_{x}-\II m_{y}\right)/\sqrt{2}\\
M_{1,0}^{(1)}=-m_{z}^{(1)},\: & M_{1,\pm1}^{(1)}=\left(\pm m_{x}^{(1)}-\II m_{y}^{(1)}\right)/\sqrt{2}\\
T_{1,0}=T_{z},\: & T_{1,\pm1}=\left(\mp T_{x}+\II T_{y}\right)/\sqrt{2}\\
T_{1,0}^{(1)}=-T_{z}^{(1)},\: & T_{1,\pm1}^{(1)}=\left(\pm T_{x}^{(1)}-\II T_{y}^{(1)}\right)/\sqrt{2}\\
Q_{2,0}^{(e)}=3Q_{zz}^{(e)},\: & Q_{2,\pm1}^{(e)}=\sqrt{6}\left(\mp Q_{xz}^{(e)}+\II Q_{yz}^{(e)}\right)\\
 & Q_{2,\pm2}^{(e)}=\frac{\sqrt{6}}{2}\left(Q_{xx}^{(e)}\mp\II2Q_{xy}^{(e)}-Q_{yy}^{(e)}\right)\\
Q_{2,0}^{(m)}=-\frac{3}{2}Q_{zz}^{(m)},\: & Q_{2,\pm1}^{(m)}=\sqrt{\frac{3}{2}}\left(\pm Q_{xz}^{(m)}-\II Q_{yz}^{(m)}\right)\\
 & Q_{2,\pm2}^{(m)}=\frac{\sqrt{6}}{4}\left(-Q_{xx}^{(m)}\pm\II2Q_{xy}^{(m)}+Q_{yy}^{(m)}\right)\\
Q_{2,0}^{(T)}=Q_{zz}^{(T)},\: & Q_{2,\pm1}^{(T)}=\sqrt{\frac{2}{3}}\left(\mp Q_{xz}^{(T)}+\II Q_{yz}^{(T)}\right)\\
 & Q_{2,\pm2}^{(T)}=\left(Q_{xx}^{(T)}\mp\II2Q_{xy}^{(T)}-Q_{yy}^{(T)}\right)/\sqrt{6}
\end{eqnarray*}
\begin{eqnarray*}
 & O_{3,0}^{(e)}=15O_{zzz}^{(e)}\\
 & O_{3,\pm1}^{(e)}=\mp\frac{15\sqrt{3}}{2}\left(O_{zzx}^{(e)}\pm\II O_{yyy}^{(e)}\pm\II O_{xxy}^{(e)}\right)\\
 & O_{3,\pm2}^{(e)}=-3\sqrt{\frac{15}{2}}\left(O_{zzz}^{(e)}+2O_{yyz}^{(e)}\pm\II2O_{xyz}^{(e)}\right)\\
 & O_{3,\pm3}^{(e)}=\mp\frac{3\sqrt{5}}{2}\left(O_{xxx}^{(e)}-3O_{yyx}^{(e)}\pm\II O_{yyy}^{(e)}\mp\II3O_{xxy}^{(e)}\right)
\end{eqnarray*}

\begin{eqnarray*}
 & O_{3,0}^{(m)}=-O_{zzz}^{(m)}/12\\
 & O_{3,\pm1}^{(m)}=\pm\left(O_{zzx}^{(m)}\pm\II O_{yyy}^{(m)}\pm\II O_{xxy}^{(m)}\right)/8\sqrt{3}\\
 & O_{3,\pm2}^{(m)}=\frac{\sqrt{2}}{8\sqrt{15}}\left(O_{zzz}^{(m)}+2O_{yyz}^{(m)}\pm\II2O_{xyz}^{(m)}\right)\\
 & O_{3,\pm3}^{(m)}=\pm\left(O_{xxx}^{(m)}-3O_{yyx}^{(m)}\pm\II O_{yyy}^{(m)}\mp\II3O_{xxy}^{(m)}\right)/24\sqrt{5}
\end{eqnarray*}

\subsection*{Material constants used in simulations}

The constants of the materials used for simulations have been measured
and provided by Sandia National Laboratories (US) in a private communication.
An infrared variable angle spectroscopic ellipsometer (J. A. Woolam)
was used to measure $\Psi$ and $\Delta$, from which the optical
constants were derived. The same constants have been used to model
the response of the previously demonstrated 3D cubic metamaterial
based on SAMPL technology \cite{sandia10}.

\begin{figure}
\includegraphics{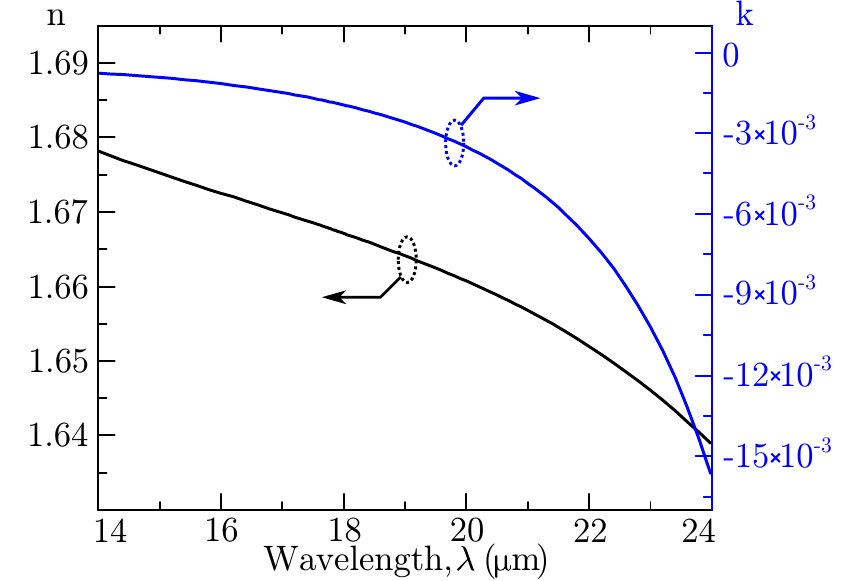}

\caption{\textbf{The refractive index of the polymer SU8 used for simulating
the case-study metamaterial.}  The complex-valued refractive index
of SU8 is given by $\tilde{n}=n+\II k$.}

\label{fig:Su8RefIndex}
\end{figure}

\begin{figure}
\includegraphics{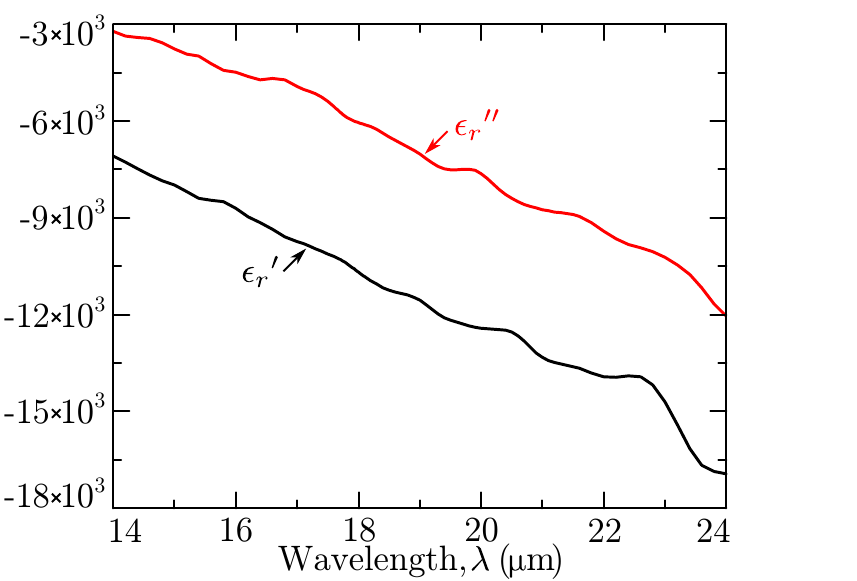}

\caption{\textbf{The dielectric constant of the gold used for simulating the
case-study metamaterial. }The complex-valued dielectric constant is
given by $\tilde{\epsilon}_{r}=\epsilon_{r}'+\II\epsilon_{r}''$ }

\label{fig:AuEr}
\end{figure}

The refractive index of the SU8 polymer that housed the gold split
ring resonators ($\tilde{n}=n+\II k$; negative $k$ implies losses)
is shown in Fig.~\ref{fig:Su8RefIndex}. The dielectric constant
of the gold used for simulations is shown in Fig.~\ref{fig:AuEr}
($\tilde{\epsilon}_{r}=\epsilon_{r}'+\II\epsilon_{r}''$; negative
$\epsilon_{r}''$ implies losses).

\renewcommand{\bibnumfmt}[1]{{#1}.{\,}}
\renewcommand\refname{References}


\end{document}